\documentclass[prd,10pt,aps,double-spaced,showkeys,twocolumn,superscriptaddress]{revtex4-1}
\usepackage{graphicx,float,amsmath,bbm,bm,array,subfigure}
\usepackage{mathtools,tabularx,xcolor,colortbl}
\usepackage[linktocpage=true,colorlinks=true,linkcolor=blue,citecolor=blue]{hyperref}
\graphicspath{{./graphics/}}
\def\nn{\nonumber\\}

\def\be{\begin{equation}}
\def\ee{\end{equation}}
\def\bearr{\begin{eqnarray}}
\def\eearr{\end{eqnarray}}

\begin{document}
	
	\title {Effect of finite volume on thermodynamics of quark-hadron matter}

	\author{Somenath~Pal}
	\affiliation{
		School of Physical Sciences,  National Institute of Science Education and Research Bhubaneswar,  HBNI,  Jatni 752050, Odisha, India}
  \affiliation{Variable Energy Cyclotron Centre, 1/AF, Bidhan Nagar , Kolkata-700064, India}

	\author{Anton~Motornenko}
	\affiliation{Frankfurt Institute for Advanced Studies, Ruth-Moufang-Str. 1, D-60438 Frankfurt am Main, Germany}

	\author{Volodymyr Vovchenko}
	\affiliation{Physics Department, University of Houston, Box 351550, Houston, TX 77204, USA}
    \affiliation{Frankfurt Institute for Advanced Studies, Ruth-Moufang-Str. 1, D-60438 Frankfurt am Main, Germany}
	\author{Abhijit~Bhattacharyya}
	\affiliation{
		Department of Physics,
		University of Calcutta,
		92, A.P.C. Road, Kolkata-700009, India}
	\author{Jan~Steinheimer}
	\affiliation{Frankfurt Institute for Advanced Studies, Ruth-Moufang-Str. 1, D-60438 Frankfurt am Main, Germany}
	\author{Horst~Stoecker}
	\affiliation{
		Institut f\"ur Theoretische Physik,
		Goethe Universit\"at, D-60438 Frankfurt am Main, Germany}
	\affiliation{Frankfurt Institute for Advanced Studies, Ruth-Moufang-Str. 1, D-60438 Frankfurt am Main, Germany}
	\affiliation{
		GSI Helmholtzzentrum f\"ur Schwerionenforschung GmbH, D-64291 Darmstadt, Germany}

\begin{abstract}
    The effects of a finite system volume on thermodynamic quantities, such as the pressure, energy density, specific heat, speed of sound, conserved charge susceptibilities and correlations, in hot and dense strongly interacting matter are studied within the parity-doublet Chiral Mean Field (CMF) model. 
    Such an investigation is motivated by relativistic heavy-ion collisions, which create a blob of hot QCD matter of a finite volume, consisting of strongly interacting hadrons and potentially deconfined quarks and gluons. 
    The effect of the finite volume of the system is incorporated by introducing a lower momentum cut-offs in the momentum integrals appearing in the model, the numerical value of the momentum cut-off being related to the de Broglie wavelength of the given particle species. 
    It is found that some of these quantities show a significant volume dependence, in particular those sensitive to pion degrees of freedom, and the crossover transition is generally observed to become smoother in finite volume. 
    These findings are relevant for the effective equation of state used in fluid dynamical simulations of heavy-ion collisions and efforts to extract the freeze out properties of heavy-ion collisions with susceptibilities involving electric charge and strangeness.	
\end{abstract}
\maketitle

\section{Introduction}
The study of the properties of strongly interacting matter under extreme conditions of temperature and density is an active area of research for some decades. It is expected that strongly interacting matter should show a phase transition between confined hadronic degrees of freedom and deconfined quark-gluon plasma (QGP) phase. Experimental programs like the Large Hadron Collider (LHC) at CERN and Relativistic Heavy-ion Collider (RHIC) at BNL have enriched our understanding of such states to a great extent. Future facilities like FAIR at GSI will join their hand in such explorations in the near future.

The matter formed in relativistic heavy-ion collisions has a finite volume. Therefore, for a comparison of the experimental findings with theoretical insights, the effects of finite volume must be taken into account. The absence of a well-defined volume of such systems and its smallness can lead to non-equilibrium effects which are beyond the scope of effective models based on thermodynamic equilibrium. On the other hand, models which describe the dynamics of heavy-ion collisions on the basis of locally equilibrated thermal systems e.g. fluid dynamic models~\cite{Hirano:2005wx,Nonaka:2006yn,Kolb:2000fha,Kolb:2001qz,Huovinen:2001cy,Hirano:2001eu,Kolb:2003dz,Petersen:2008dd,Gale:2012rq} have been very successful in describing bulk matter properties.

Finite volume effects in heavy-ion collisions depend on the size of the colliding nuclei, the center of mass energy, $\sqrt{s_{\rm NN}}$, and centrality of collisions as well as the size of the regions of homogeneity which can be considered coherent thermal systems. There have been some studies regarding the values of the volume of such systems both theoretically and experimentally. HBT radii measurements~\cite{CERES:2002rfr} imply that the freeze-out volume increases with the increase of $\sqrt{s_{\rm NN}}$. The freeze-out volumes have been analyzed by the ALICE Collaboration~\cite{ALICE:2014xrc} and found radii as large as 10 fm. Within the Ultrarelativistic Quantum Molecular Dynamics (UrQMD) model~\cite{Bleicher:1999xi} framework, the volume of homogeneity has been calculated in Ref.~\cite{Graef:2012sh} and has been compared with experimental results. The volume of homogeneity with transverse momentum window of 300-400 MeV was found to vary from 50 to 250~fm$^3$ in the case of lead-lead collision at different centralities for  $\sqrt{s_{\rm NN}}$ in the range of 62.4 to 2760 GeV. It is expected that the system size is even smaller during the initial equilibration time~\cite{Bozek:2013df,Bzdak:2013zma}. It is therefore relevant to investigate finite size effects on a system with a radius less than 10 $fm$ within an effective model approach. 

Finite size scaling analysis~\cite{Ferdinand:1969zz,Fisher:1972zza} is a tool to understand the effects of finite size on the properties of a system of strongly interacting matter. Such analyses have been considered in Ref.~\cite{Abreu:2006pt,Palhares:2009tf,Fraga:2011hi,Almasi:2016zqf}. It was found that there is a smoothing and broadening of singularities in a finite volume, which is expected since it is well known that in a finite volume no strict first order phase transition can exist and infinite correlation lengths are not possible. The T-$\mu_B$ region, where discontinuities in the chiral condensate are observed, is shrunken and the pseudo-critical lines are shifted to a higher chemical potential region. 
Other theoretical studies regarding finite volume have been done in Ref.~\cite{Elze:1986db,Spieles:1997ab,Gopie:1998qn,Bazavov:2007zz,Fischer:2005nf,Luecker:2009bs,Luscher:1985dn,Gasser:1987ah,ELZE1986385,PhysRevC.68.044907,PhysRevC.57.908}. These works also indicate smoothing of the singular behavior of the system.

The effects of finite sizes in QCD matter have been subject of discussion in different contexts. 
For example in Ref.~\cite{Neergaard:1998xg}, the authors have considered the effect of the surface term in the free energy of a bubble of quark or hadron matter at the quark-hadron phase transition. Such an approach is relevant in the vicinity of a first order phase transition and valid for systems that form spherical bubbles of finite sizes. Similar problems were investigated in \cite{Ke:2013wga} where interface effects such as the interface tension, the interface entropy and critical bubble size where studied in the NJL model and in \cite{Kroff:2014qxa} where the nucleation of quark matter in magnetars and the effect of finite droplets was discussed. In \cite{Xia:2018wdj} the surface tension and curvature term of strange quark matter was investigated.
In numerical lattice simulations artifacts can arise from the finiteness of the practical lattice size used which was discussed in ~\cite{deForcrand:2005pb} in the context of the dual string tension of a spatial ’t Hooft loop in the deconfined phase of Yang-Mills theory.

In this paper, we discuss mainly the properties of QCD matter at vanishing chemical potentials where no phase coexistence occurs and thus no clear interface between two phases. Instead, we will use a simple method to calculate the thermodynamic quantities of  strongly interacting matter with finite size by introducing a lower momentum cut-off. This method assumes that the total force on a particle exerted by all the other particles in the finite system is position-independent, as long as the test particle is inside the finite volume of our interest. 
Here, excluding the states with momentum less than the cut-off mimics the lowering of the energy density due to the limited number of particles in the small system. Therefore, this simplified approach does not exclude any term with a specific power of the system size.

The specific heat, speed of sound, and susceptibilities of conserved charges all provide useful information about the medium properties, degrees of freedom, and interactions in the QCD system \cite{Stodolsky:1995ds,Shuryak:1997yj,Basu:2016ibk,Bhattacharyya:2013oya,Bhattacharyya:2017gwt,Fukushima:2008wg,Roessner:2006xn,Sasaki:2006ww,Ratti:2007jf,Fukushima:2009dx,Schaefer:2006ds,Schaefer:2007pw,Schaefer:2009ui,Borsanyi:2018grb,Bazavov:2020bjn,Bhattacharyya:2015zka,Bhattacharyya:2014uxa,Koch:2008ia,Koch:2005vg}. 
The purpose of this paper is to investigate the effect of a finite volume on the above thermodynamic quantities within the framework of a Chiral Mean Field model formulated in the grand canonical ensemble. 

The paper is organized as follows: In section (II) we discuss the Chiral Mean Field model briefly and introduce the finite size corrections. The Results are presented and discussed in section (III). In section (IV) we conclude and make remarks on our findings.

\section{Chiral Mean Field model}
To investigate the effect of a finite size in heavy-ion collisions we will employ the Chiral Mean Field model (CMF) which has been developed over the past years. The reason for this choice is that the CMF contains both a hadronic as well as deconfined phase with a smooth crossover consistent with lattice~QCD thermodynamics at $\mu_B=0$. To not be overwhelmed with all the gory details of the CMF we refer the reader to~\cite{Papazoglou:1998vr,Steinheimer:2010ib,Motornenko:2019arp} for a detailed description of the model and will in the following only recapitulate the most important aspects.
The CMF is a chiral SU (3)-flavor parity-doublet Polyakov-loop quark-hadron mean-field model and provides an effective approach to describe the interplay of hadronic and deconfined QGP states, i.e. the phase diagram of strongly interacting matter. In the hadronic phase, the baryonic octet and all its parity partners interact with the chiral mean-field.

In this scenario the effective masses of the ground state octet baryons and their parity partners (assuming isospin symmetry) read~\cite{Steinheimer:2011ea}:
\begin{eqnarray}
m^*_{b\pm} = \sqrt{ \left[ (g^{(1)}_{\sigma b} \sigma + g^{(1)}_{\zeta b}  \zeta )^2 + (m_0+n_s m_s)^2 \right]} \nn\ \pm g^{(2)}_{\sigma b} \sigma \ ,
\end{eqnarray}
where the various coupling constants $g^{(*)}_{*b}$ are chosen such that they reproduce the vacuum masses and nuclear matter properties, $n_s$ is the SU(3) breaking mass term that generates an explicit mass corresponding to the strangeness. The above mass formula shows a mass splitting between the baryon parity partners which is generated by the scalar mesonic fields $\sigma$ and $\zeta$~\cite{Detar:1988kn,Zschiesche:2006zj,Aarts:2017rrl,Sasaki:2017glk}.

Here, $\omega$ is the field responsible for vector repulsion at finite baryon densities, $\rho$ is the repulsive vector field at finite isospin densities and $\phi$ is for vector repulsion at finite strangeness densities.

In addition, all the hadrons listed in Particle Data Tables~\cite{Tanabashi:2018oca} contribute to the system as multi-component hadron resonance gas.

The coupling constants and the parameters of the effective potential for these fields \cite{Steinheimer:2011ea} are described in much more detail in~\cite{Motornenko:2020yme}.

To include deconfinement and quark degrees of freedom, a Polyakov-loop-extended Nambu-Jona-Lasinio (PNJL) inspired model~\cite{Fukushima:2003fw} is used.

For the quarks, the $\sigma$ and $\zeta$ fields dynamically generate the masses $m_q^*$ of the quarks.
\begin{eqnarray}
m_{u,d}^*  &=&-g_{l\sigma}\sigma+\delta m_l + m_{0}\nonumber\\
m_{s}^*  &=&-g_{s\zeta}\zeta+ \delta m_s + m_{0}\
\end{eqnarray}
$m_{0q}$ is the contribution to the quark mass from gluon condensate.

The phase transition is controlled by two mechanisms: the quarks appear when the Polyakov loop order parameter becomes finite, and hadrons become suppressed due to their excluded volumes as the system's pressure and density increase.
Assigning an excluded volume to the hadrons suppresses their number density in the system at high pressure~\cite{Rischke:1991ke,Steinheimer:2011ea}. We take the excluded volume of all the mesons $v_M = 1/8$~fm$^3$ and that of baryons $v_B=1$~fm$^3$. 
In principle already the inclusion of a finite size of the hadrons, preventing the hard cores of hadrons to overlap, will already introduce finite size effects if the system size becomes of the order of the largest hadron size~\cite{Vovchenko:2018cnf}. In addition, the excluded volume effect will lead to significant modifications of the pressure and susceptibilities of the system, as studied in \cite{Vovchenko:2014pka,Vovchenko:2016rkn,Alba:2016hwx,Satarov:2016peb}. A comparative study of charge susceptibilities for various combinations of excluded volumes for infinite volume, within the CMF, has been presented in Ref.~\cite{Motornenko:2020yme}. However, it should be noted, that the excluded volume effect is different from the finite system volume effect that we study here, where the finite size effects stem from the fact that the wave function of hadrons with small momenta cannot be localized in a finite total volume. Therefore, in the following, we will study the finite size effects assuming a fixed set of excluded volumes and under the assumption that the total volume is always significantly larger than the largest excluded volume $v_B$ (see e.g. \cite{Vovchenko:2018cnf,Poberezhnyuk:2020ayn} for a discussion what happens as one approaches that limit). Consistent treatment of small systems which are of similar size as the eigenvolume of the largest hadrons is out of the scope of the present work.

To untangle the various contributions from the confined vs. deconfined part of the system as well as understand the effect of the excluded volume, we will also compare the full CMF results with a reduced model which contains only the hadronic part of CMF and neglects the excluded volume effect (keeping all other parameters fixed). In such a scenario, which we refer to as \textit{pure hadronic CMF}, the hadronic degrees of freedom to are not suppressed by the excluded volume term and thus numerously appear at increasing temperature, with the excitation of an increasing number of heavier and heavier resonances. Historically, such an effect has led to the discovery of the ``limiting'' Hagedorn temperature \cite{Hagedorn:1965st}, which can be tamed with the excluded volume effect~\cite{Gorenstein:1981fa,Vovchenko:2018eod} and which is now understood to signal the liberation of partonic degrees of freedom. 
In the following, we do not focus on the physics of Hagedorn temperature but rather discuss how different contributions from hadrons vs. quarks to the finite volume effects can be separated. One should also note that the \textit{pure hadronic CMF} still goes beyond a simple hadron resonance gas model, as the hadronic degrees of freedom interact with the chiral mean-fields of the model leading to chiral restoration in the high temperature and density part of the phase diagram.

The effect of finite volume is incorporated by introducing a non-zero lower momentum cut-off~\cite{Bhattacharyya:2015zka}
\begin{equation}
 p_{min}=\pi/R=\lambda   
\end{equation}
in all the momentum integrations, where R is the diameter of the spherical volume $V=\frac{4}{3}\pi R^3$. This constraint on $p_{min}$ can be obtained directly from the fact that the maximum de-Broglie wavelength of a particle inside a finite volume is $2R=\frac{h}{p}$. The argument behind this is that the de-Broglie wavelength of any particle cannot be larger than twice the size of the system.  Such a consideration excludes the low-lying energy levels, reducing the number density of particles and hence,  reducing the pressure as compared to that of an infinitely large system with the same temperature and chemical potentials.




Such an approach has been applied before on HRG model and PNJL model separately where it was found that the properties of the system depend significantly on its size~\cite{Bhattacharyya:2012rp,Bhattacharyya:2015zka,Bhattacharyya:2014uxa,Redlich:2016vvb,Saha:2017xjq,Palhares:2011jf,Tripolt:2013zfa,Bernhardt:2022mnx,Kovacs:2023kcn}. The restriction of the system to a finite volume leads to discrete energy levels available for the constituting particles. But for simplicity, we take integration over particle momenta instead of summation. This approximation has been shown to be accurate~\cite{Karsch:2015zna}.
Alternatively, one could consider a system with (anti)periodic boundary conditions, which would resemble better the systems studied with lattice gauge theory. In this case, one would have to pay special attention to zero momentum modes that appear in such systems~\cite{Bernhardt:2021iql}.
The lower momentum cutoff studied here corresponds to a system where the matter vanishes outside the finite volume which seems appropriate for heavy-ion collisions.

This lower momentum cut-off approach to mimic the effect of finite volume has been previously used to study the effect of finite volume in the Polyakov loop extended Nambu-Jona-Lasinio (PNJL) model~\cite{Bhattacharyya:2015kda,Saha:2017xjq}. It was found that the ratios of susceptibilities depend on system volume even in the absence of hadrons. 

We have checked that the approach of using a momentum cutoff preserves thermodynamic consistency, similar to conclusions found in \cite{Redlich:2016vvb}  and also neglect surface and curvature~\cite{SomorendroSingh:2008fa} effects which would be more relevant in the presence of a phase transition and the formation of droplets.

\section{Results and discussion}

\subsection{Thermodynamic quantities}
In this section we present the results for several thermodynamic quantities, including scaled pressure $(P/T^4)$, scaled energy density $(\epsilon/T^4)$, specific heat at constant volume $(C_V)$, speed of sound squared $(c_s^2)$, ratios of susceptibilities and correlations of conserved charges for different system volume sizes at zero chemical potentials. 

\begin{figure}[t]
	\begin{center}
			\includegraphics[width=8.5cm,height=6cm]{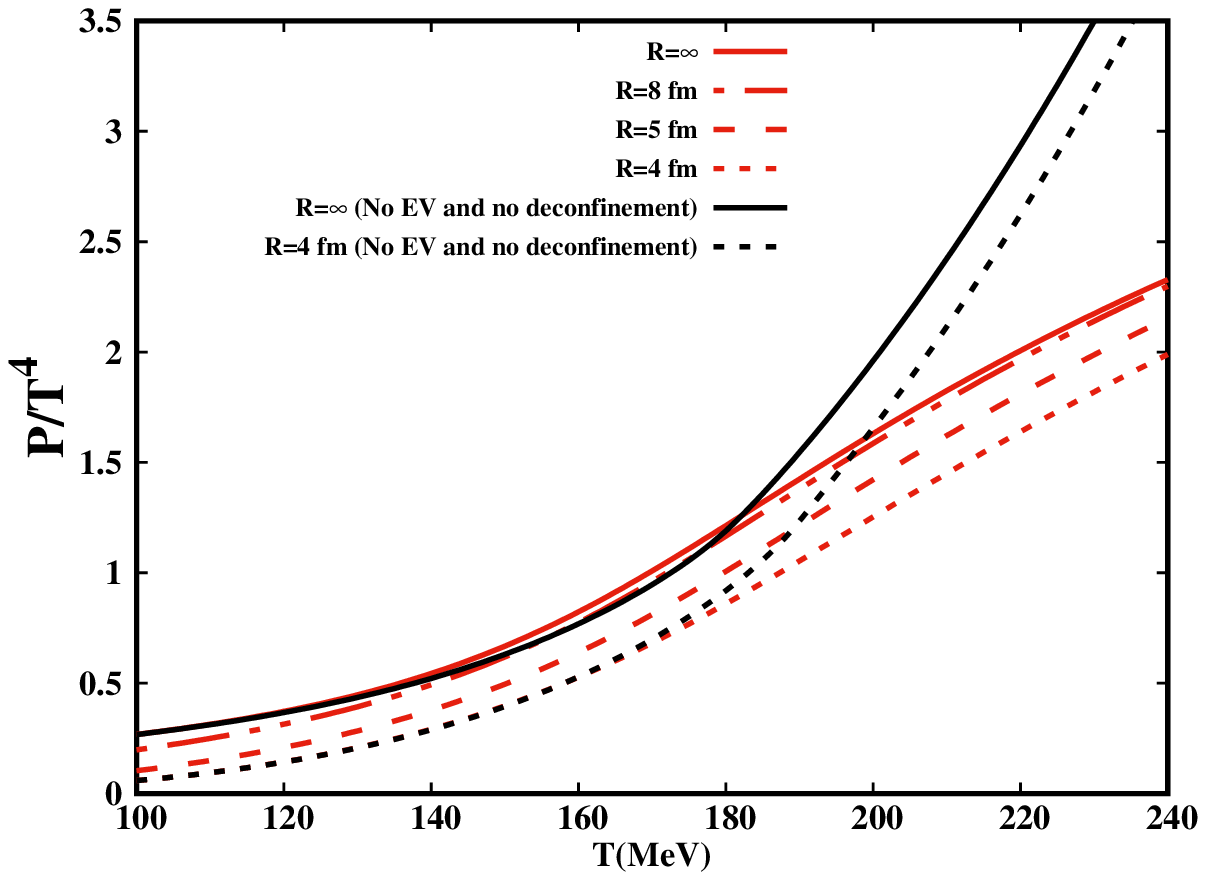}\\
			\includegraphics[width=8.5cm,height=6cm]{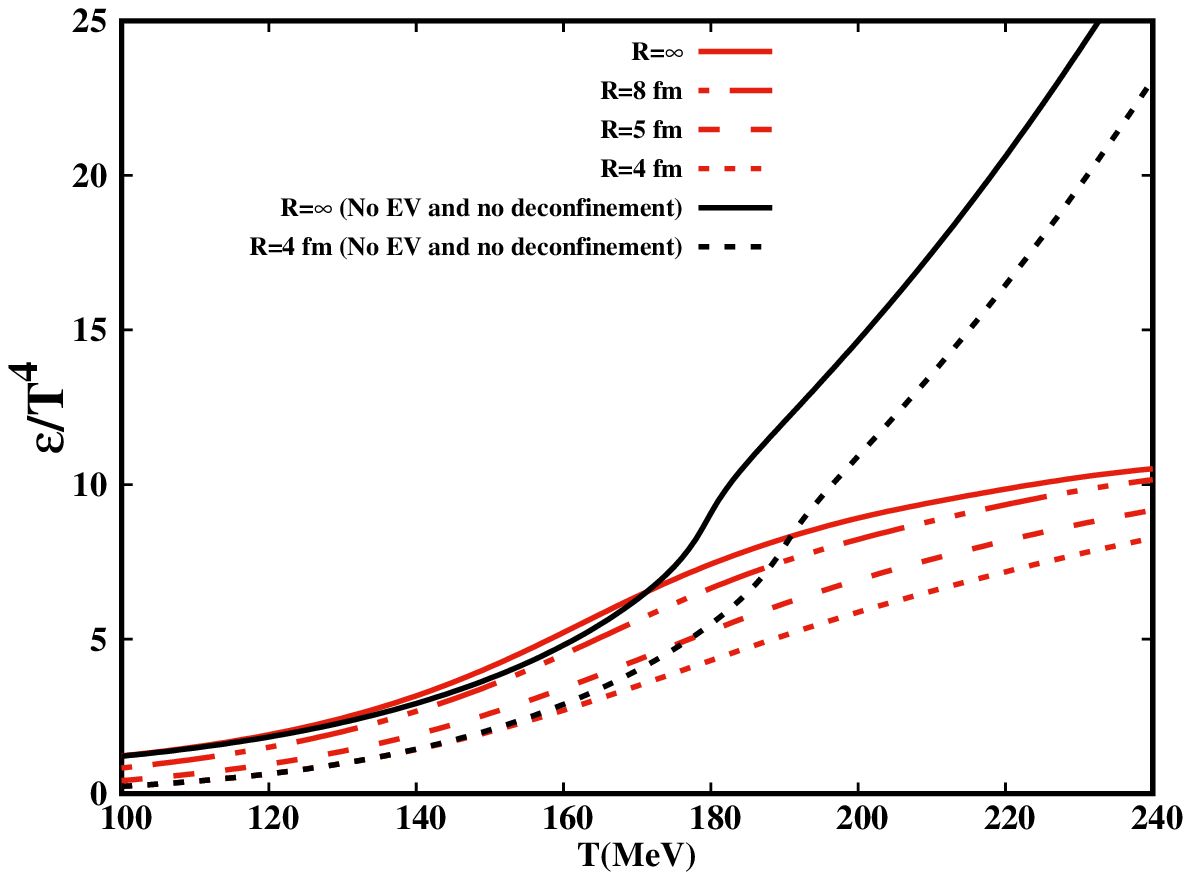}\\
		\caption{(Color online) Scaled pressure (upper figure) and scaled energy density (lower figure), in the full CMF model (red lines) and purely hadronic model (black lines), as functions of the temperature at $\mu_B=\mu_Q=\mu_S=0$. Different volume sizes are compared. Smaller system sizes lead to systematically smaller pressure and energy densities.}
		\label{fig 1}
	\end{center}
\end{figure}

In Fig.~\ref{fig 1}, we show the dependence of the scaled pressure $(P/T^4)$ and scaled energy density $(\epsilon/T^4)$ on the temperature (at $\mu_B=\mu_Q=\mu_S=0$) for various system volumes. In the following, the red lines correspond to the full CMF model including deconfined quarks and the black lines are the purely hadronic model without excluded volume. In both cases, it is seen that the scaled pressure and scaled energy density decrease as the volume of the system is decreased.
At lower temperatures the pressure and energy density are dominated by pions, especially those with rather low momenta. For this reason, a significant dependence of the scaled pressure on the volume is observed. As the temperature increases, other, more massive degrees of freedom start to contribute more. Furthermore the average momentum is increased which means that the low-momentum cutoff removes a relatively smaller fraction of the total pressure contribution, even though the absolute shift is still significant. 

When the temperature is increased to even higher values, low mass quarks become the dominant degrees of freedom in the CMF model and the effect of the cutoff decreases. In the purely hadronic CMF model the pressure keeps increasing due to the large number of hadronic resonances available, and the effect of the momentum cutoff remains significant. 
In addition at this temperature regime the Polyakov loop potential attains significantly large values which are not affected by the momentum cut-off. The interplay of these effects (mass, excluded volume, and the Polyakov loop) govern the difference between the results of different finite volumes. 

Note that the above described systematics also lead to a larger quark number fraction at smaller system volumes. The energy density generally shows a stronger dependence on system volume than the pressure and the chiral transition at ${\mu=0}$ continues to remain a crossover although the energy density plot shows that it is smoothened to some extent for finite volume.

\begin{figure}[t]
	\begin{center}
			\includegraphics[width=8.5cm,height=6cm]{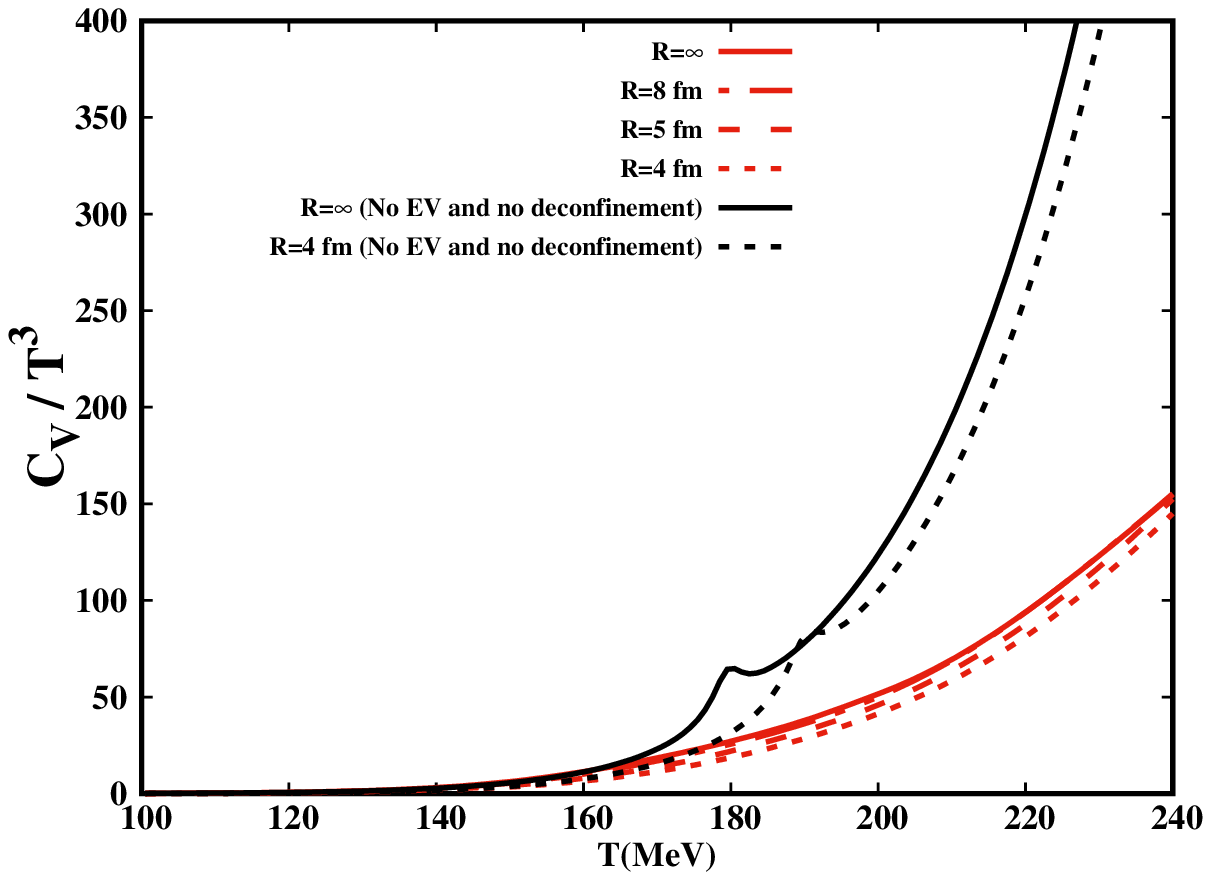}\\
			\includegraphics[width=8.5cm,height=6cm]{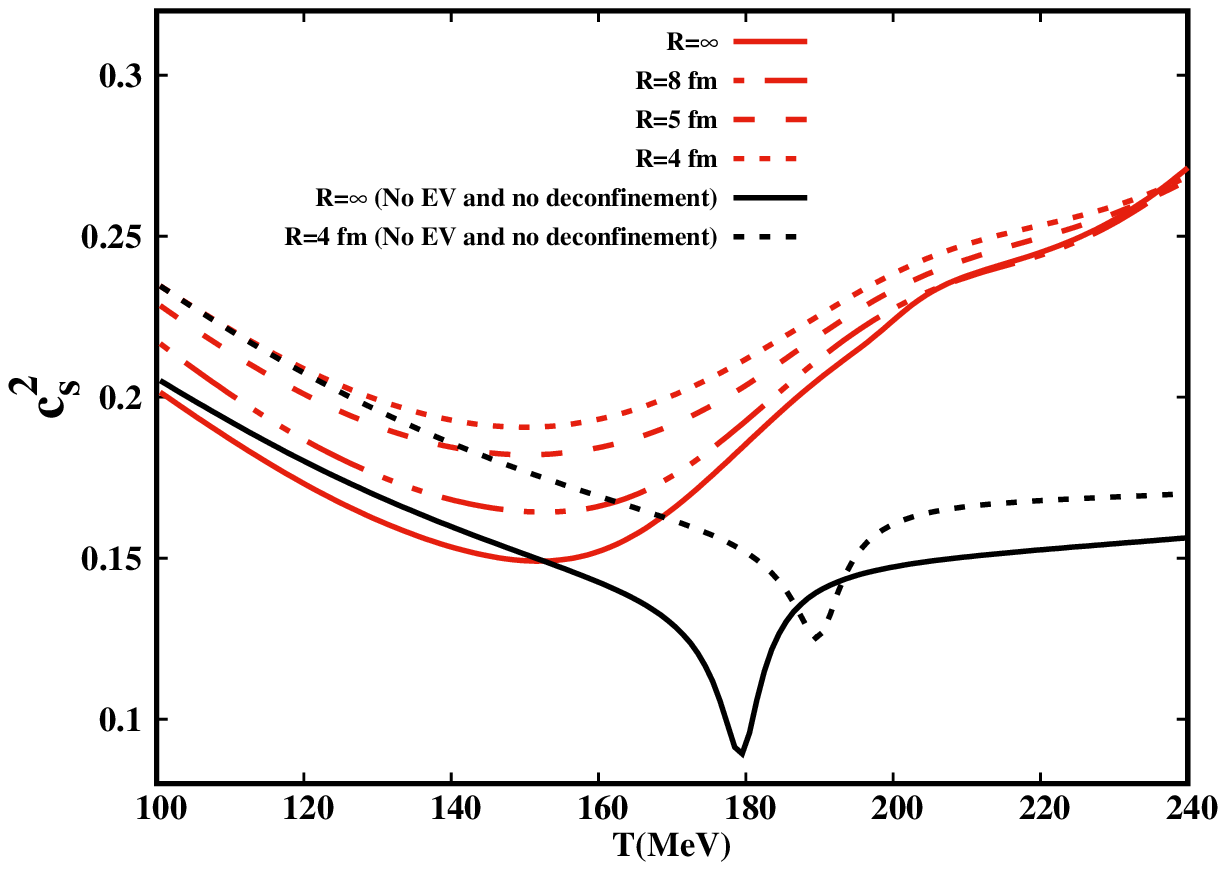}\\
		\caption{(Color online) Scaled specific heat at constant volume and speed of sound squared as functions of temperature at $\mu_B=\mu_Q=\mu_S=0$ in the full CMF model (red lines) and purely hadronic model (black lines). While the effect of a finite system size on the specific heat is only modest, the speed of sound is significantly enhanced.}
		\label{fig 2}
	\end{center}
\end{figure}

Next, we will turn to derivatives of the pressure at vanishing chemical potentials. The upper panel of Fig.~\ref{fig 2} shows that the specific heat 
\begin{equation}
  {\left.\frac{C_V}{T^3}\right|_{\mu=0}}=\frac{1}{T^3}\left(\frac{\partial \epsilon}{\partial T}\right)_{V,\mu=0} 
\end{equation}
in the CMF model with quarks increases monotonically with temperature. $C_V$ rises rapidly. In the scenario without quarks and excluded volume, the chiral crossover transition is sharper and a maximum in the specific heat appears. This is expected since the specific heat is discontinuous near a phase transition of first order and becomes singular at a phase transition of second order. In this case also a smaller system volume leads to an increased temperature of the inflection point of the pressure. Away from the crossover, the $C_V$ at a particular temperature decreases with system volume because statistical fluctuations scale as $\sqrt{N}$ where $N$=number of particles.

At low temperatures, the difference in $C_V$ for different system volumes is smaller due to smaller number of particles in the system. As the temperature rises to a medium range, this difference is the largest. At even higher temperatures, point-size quarks become the dominant particles and hence the effect of finite volume decreases.

The speed of sound is a very important property of strongly interacting matter as it gives important information that can be tested e.g. in fluid dynamic simulations of heavy-ion collisions. It can be calculated as 
\begin{equation}
    c_s^2=\Big (\frac{\partial P}{\partial \epsilon}\Big )_{\mu=0}=\Big (\frac{s}{C_V}\Big )_{\mu=0}
\end{equation}
where $s$ is the entropy density.
A pronounced minimum in the speed of sound is expected near a phase transition ~\cite{NoronhaHostler:2008ju,Castorina:2009de,Cleymans:2011fx,Gavai:1985xu,Redlich:1985uw,BraunMunzinger:1996mq} while a more shallow minimum indicates a broad crossover~\cite{Chojnacki:2007jc}. It is seen from the lower panel of Fig.~\ref{fig 2} that speed of sound squared ($c_s^2$) decreases with increasing temperatures at moderate temperatures, 
$T \lesssim 150$~MeV.
This behavior is known for a hadron resonance gas (HRG) where various resonance particles populate the system significantly and account for the attractive interaction among the ground state hadrons leading to a sluggish increment in the number of effective free degrees of freedom. Hence, the entropy density increases slowly compared to $C_V$, and $c_s^2$ decreases with increasing temperature at low temperatures.
At smaller volumes the population of light mesons is suppressed more than that of heavy baryons due to lower momentum cut-off, which should result in a reduction in entropy density with system volume. But the suppression of $C_V$ at smaller volumes is much larger than the corresponding decrease in entropy density. As a result, we see that $c_s^2$ increases when system volume is decreased.
At higher temperatures, $T \gtrsim 150$~MeV, $c_s^2$ rises when the quark degrees of freedom start to appear. 
Note that a minimum in $c_s^2$ is also present in the scenario without quarks, albeit at a higher temperature, $T \sim 180-190$~MeV.
This is in contrast to the ideal HRG model, where the speed of sound keeps decreasing at high temperatures~\cite{Vovchenko:2016rkn} and indicates a strong chiral crossover transition in the purely hadronic model without EV. The location of this crossover is also modified by the finite system size and a smaller system volume leads to a slightly higher pseudo-critical temperature.
The rise of the speed of sound at high temperatures that we observe here is due to the presence of hadronic interactions in the hadronic part of the CMF model.
As in the scenario with quarks, the finite volume effect becomes less significant at larger temperatures.

\subsection{Fluctuations of conserved charges}

The susceptibilities of conserved charges can be a very useful tool to understand the underlying degrees of freedom as well as to extend our knowledge of QCD thermodynamics into the finite density regime. These susceptibilities, $\chi_{ijk}^{BQS}$, are calculated as derivatives of the thermodynamic pressure with respect to the different chemical potentials, corresponding to the different conserved charges, baryon number, electric charge, and strangeness,
\begin{equation}
\chi^{BQS}_{ijk} = \frac{\partial^i \partial^j \partial^k P(T,\mu_B,\mu_Q,\mu_S)/T^4}{\partial \left({\mu_B/T}\right)^i \partial \left({\mu_Q/T}\right)^j \partial \left({\mu_S/T}\right)^k}\,.\
\end{equation}

The susceptibilities are not only related to the multiplicity distributions, and thus the fluctuations and correlations of the given charges, but they also define the Taylor series expansion of the scaled pressure $(P/T^4)$ with respect to the baryon, electric, and strange chemical potentials~\cite{Allton:2002zi}:
\begin{equation}
P = P_0 + T^4 \sum_{i,j,k} \frac{1}{i!j!k!} \chi_{ijk}^{BQS} \left(\frac{\mu_B}{T}\right)^i
\left(\frac{\mu_Q}{T}\right)^j
\left(\frac{\mu_S}{T}\right)^k\
\end{equation}
where $P_0$ is the pressure at chemical potential=0.

\begin{figure}[t]
	\begin{center}
			\includegraphics[width=0.48\textwidth]{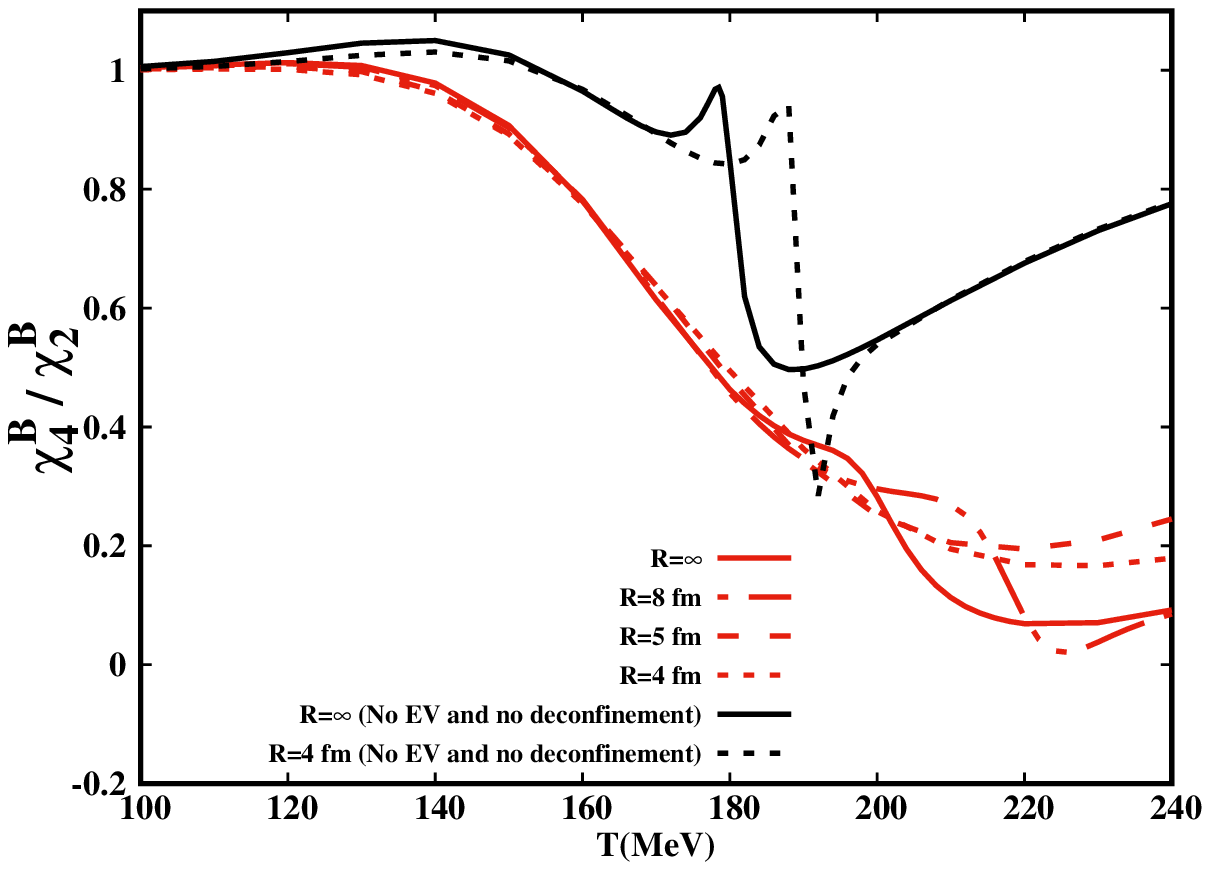}\\
			\includegraphics[width=0.48\textwidth]{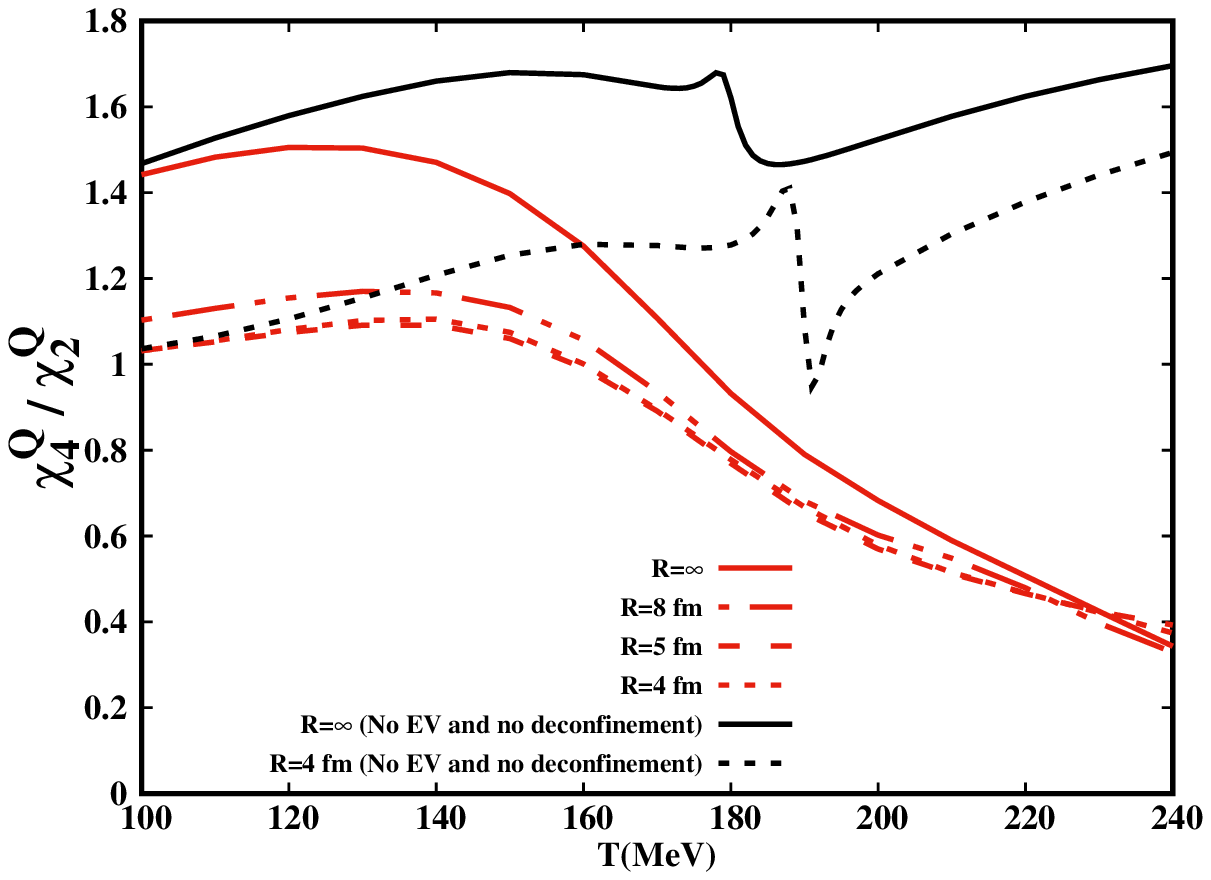}\\
			\includegraphics[width=0.48\textwidth]{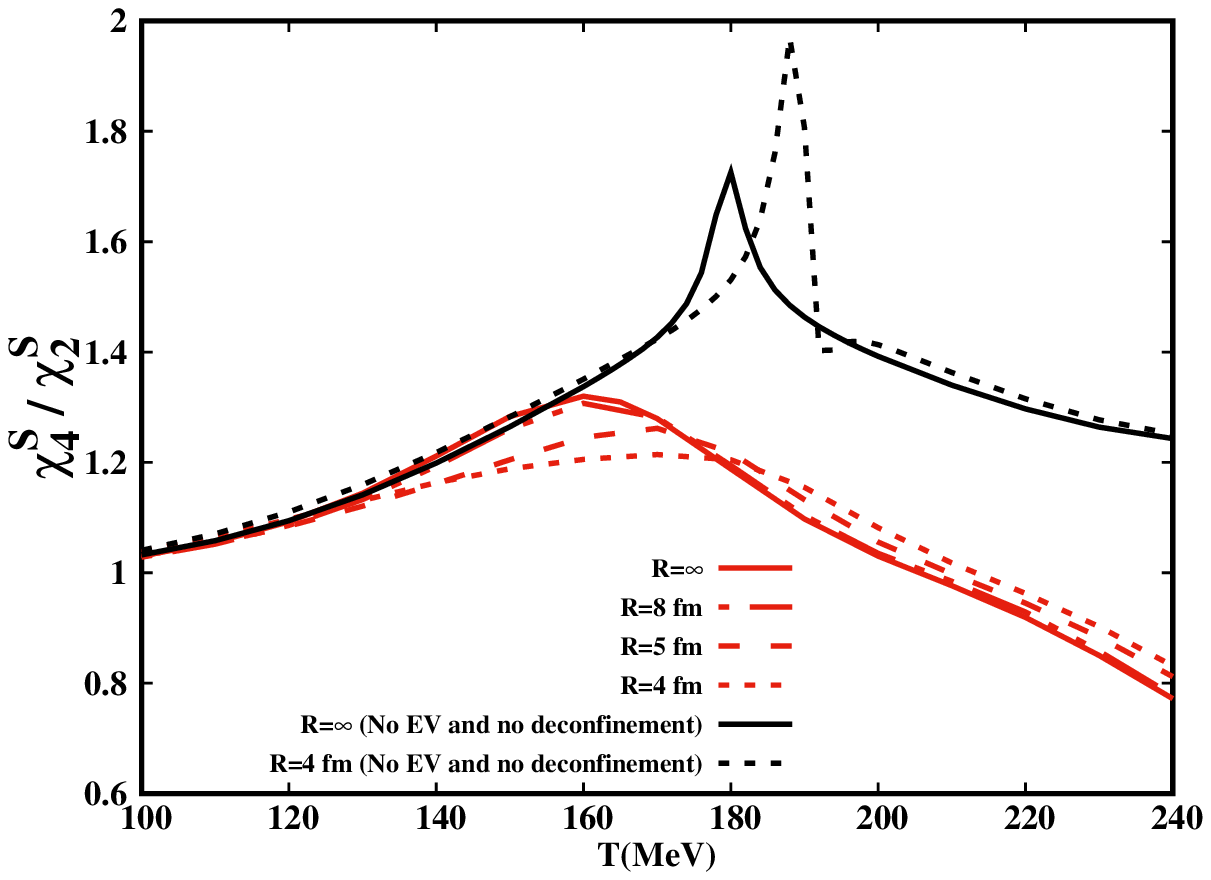}
		\caption{(Color online) Ratios of fourth to second order susceptibilities of different conserved charges as function of the temperature at $\mu_B=\mu_Q=\mu_S=0$ in the full CMF model (red lines) and purely hadronic model (black lines).
		}
		\label{fig 5}
	\end{center}
\end{figure}

It has been argued that direct measurements of the different susceptibility ratios in heavy-ion collisions may allow the determination of the hadronization temperature \cite{Alba:2014eba,Bellwied:2013cta}, QCD critical point~\cite{Stephanov:1999zu,Bzdak:2019pkr}, or even the speed of sound \cite{Sorensen:2021zme} of hot QCD matter.
Attempts at directly extracting these susceptibilities from experiment and relating them to system properties are already complicated due to the finite size, finite lifetime, canonical ensemble effects, and complex freeze-out dynamics of the system~\cite{Vovchenko:2021gas}. 
While for sufficiently large volumes ratios of susceptibilities have the advantage of not depending on the volume, in the following we will show that, even in thermal equilibrium, a finite size of the system, comparable to that expected in heavy-ion collisions, will have a significant impact on these fluctuation measures.

Figure~\ref{fig 5} shows the ratios of the fourth to second order susceptibilities for the baryon number, electric charge, and strangeness at $\mu_B=\mu_Q=\mu_S=0$, for different system sizes. 

First, we can observe that the baryon number susceptibilities are almost unaffected by the finite system volume. This is due to the large masses of the nucleons and their resonances, which dominate the baryonic thermodynamics in the low and intermediate mass range. Only at the highest temperatures, where the lighter quarks dominate, we start observing a difference. This is not observed in the purely hadronic model. The position of the kink in the CMF (for $T>200$ MeV) depends on the system volume and is related to the appearance of strange quarks, coupled to the chiral fields, which is delayed in temperature compared to the light up and down quarks. Also the purely hadronic model without EV exhibits a strong kink in all three susceptibility ratios. This is due to the strong chiral crossover and, as for the speed of sound, the location of these kinks depends on the location of the pseudo-critical temperature and thus on the systems volume.

\begin{figure}[t]
	\begin{center}
		\includegraphics[width=0.47\textwidth]{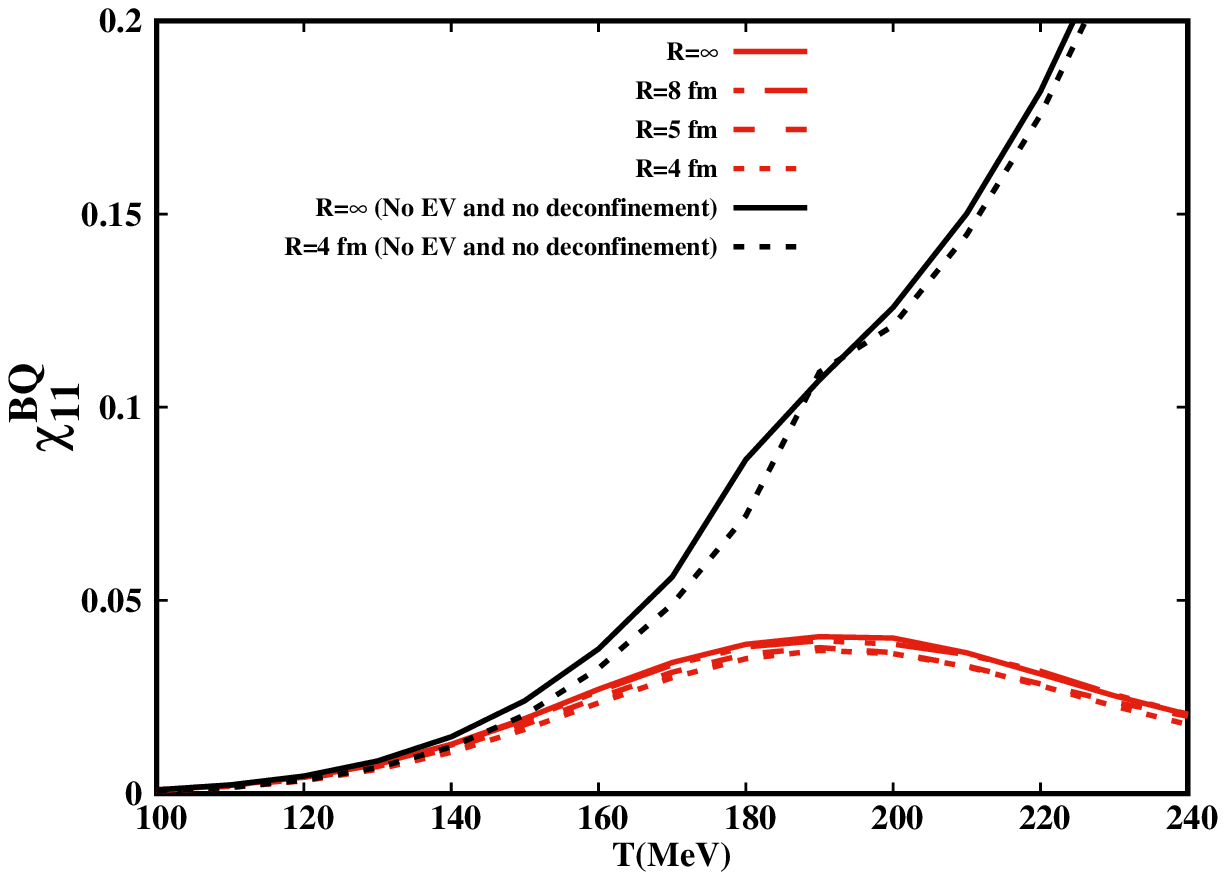}\\
		\includegraphics[width=0.47\textwidth]{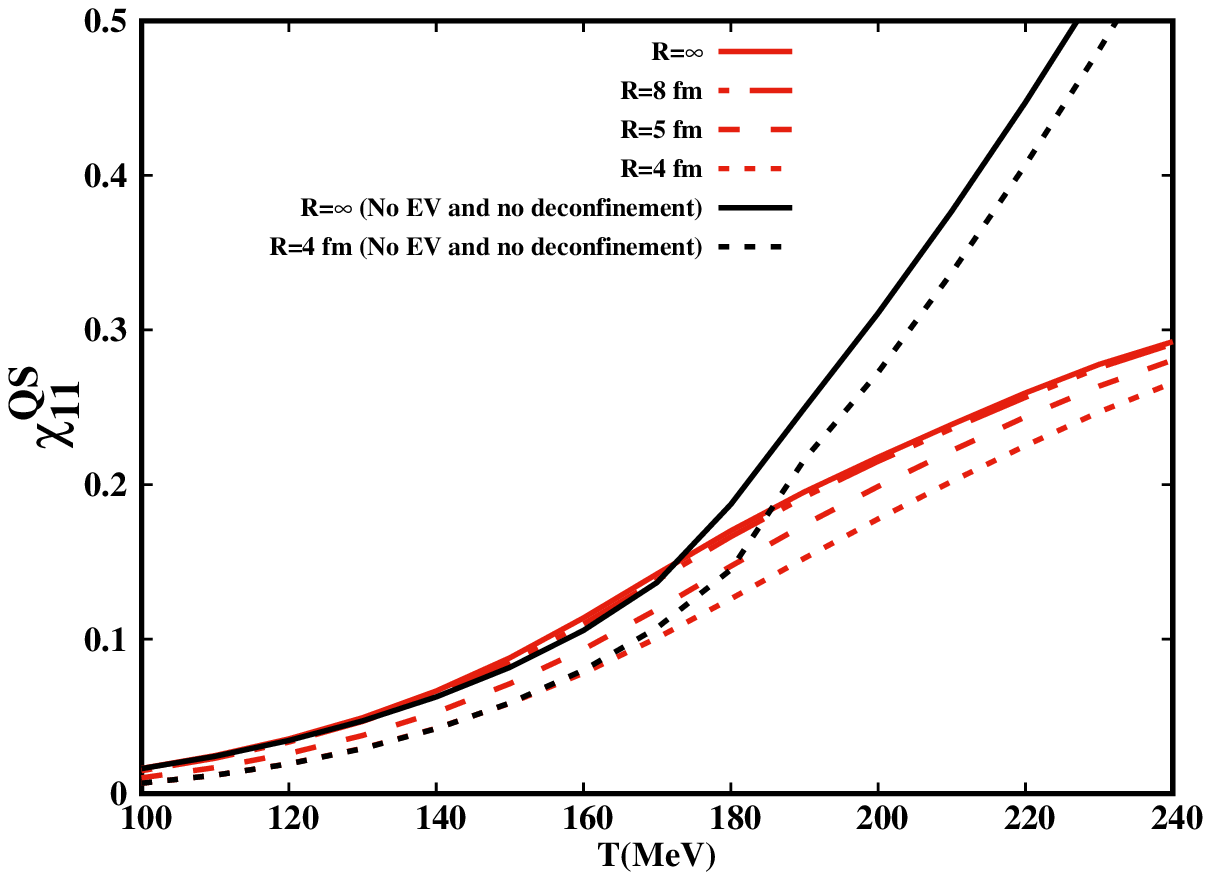}\\
		\includegraphics[width=0.47\textwidth]{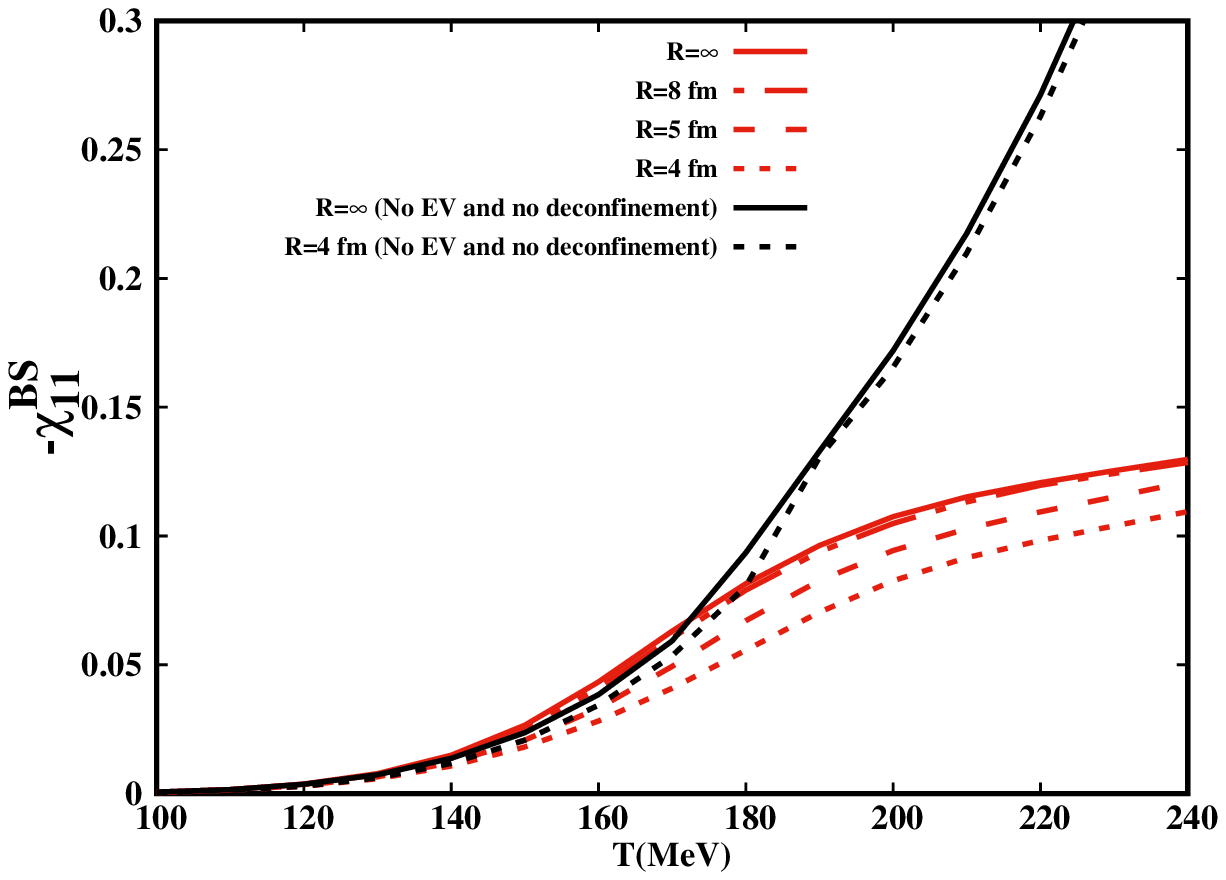}
		\caption{(Color online) First order off-diagonal susceptibilities as function of temperature at $\mu_B=\mu_Q=\mu_S=0$ for different finite volumes , in the full CMF model (red lines) and purely hadronic model (black lines). A significant effect is observed in the $QS$ and $BS$ correlations.}
		\label{fig 6}
	\end{center}
\end{figure}

The ratio of electric charge susceptibilities, on the other hand, shows a significant dependence on system volume, especially at low temperatures in the full CMF model. The dominant contributors in this sector are the pions which have a very low mass and therefore are significantly affected by the momentum cut-off as well as by Bose statistics. Already small electric chemical potentials will lead to an enhancement of low-momentum pions due to the Bose-statistics, thus a low-momentum cut-off will be more severely felt by the pions. 
At low temperatures in particular, the finite size effects lead to a strong suppression of the $\chi_4^Q/\chi_2^Q$ ratio, even for the largest finite system size considered, $R = 8$~fm. In the purely hadronic model this strong effect remains visible up to large temperatures as here the pions are not replaced by the deconfined quarks. 
This effect is not observed for the strangeness susceptibilities due to the much higher mass of even the lightest strange hadrons, the kaons. In the hadronic model, dominated by the kaons, no effect from the volume, besides small shift of the peak associated with the chiral crossover, is observed. The local maximum in the full CMF model must therefore be from the appearance of the strange quarks.
These results have relevance both for using net-charge fluctuations in analyzing freeze-out in heavy-ion collisions, as well as lattice QCD calculations that are performed in a finite volume. 
Indeed, the results show that it might be challenging to extrapolate the finite volume calculations of $\chi_4^Q/\chi_2^Q$ to the thermodynamic limit.

Fig.~\ref{fig 6} shows the off-diagonal susceptibilities related to charge correlations of baryon,  electric and strange charges at $\mu_B=\mu_Q=\mu_S=0$ for different system volumes. 

At low temperatures, the number density of baryons is small. Hence $\chi_{11}^{BQ}$ is very small. As the temperature increases, hadrons with both nonzero baryon number and electric charge begin to appear. In particular the heavy Delta baryon plays an important role in this baryon-charge correlation. In addition to that, quarks also begin to populate the system at higher temperatures, leading to saturation. When the combined contribution due to electrically charged baryons and up quarks becomes maximum, we see a peak in $\chi_{11}^{BQ}$ at $T=190$ MeV approximately. For even higher temperatures, the number of hadrons, in particular $\Delta$'s, decreases due to the excluded volume in the full CMF model. Consequently $\chi_{11}^{BQ}$ decreases at high temperatures and slowly approaches its asymptotic value of zero. In the model without quarks this does not appear and $\chi_{11}^{BQ}$ keeps increasing. 

Electrically charged strange mesons are significantly populated at low temperatures. Hence, $\chi_{11}^{QS}$ is significantly larger than $\chi_{11}^{BQ}$ at all temperatures. Strange quarks show up at somewhat higher temperatures than up and down quarks which means that this correlation is dominated by the hadronic degrees of freedom to higher temperatures. The effect of the system volume on $\chi_{11}^{QS}$ is larger than that on $\chi_{11}^{BQ}$ because the electric charge-strangeness correlation is dominated by the lighter strange mesons as compared to the heavy charged baryons dominating the baryon-electric charge correlation. Both $\chi_{11}^{BQ}$ and $\chi_{11}^{QS}$ decrease with the system volume. 

The baryon-strangeness correlation is negative at all temperatures because of the opposite sign of baryon number and strangeness of hyperons. Although it is seen in the plot that $\chi_{11}^{BS}$ for smaller system volume is larger than that for larger volume, the absolute value of $\chi_{11}^{BS}$ for larger volume is larger. The trend is again understood due to the large masses of strange baryonic hadrons which dominate this correlation. On the other hand, we observe that this correlation in the purely hadronic model does not change with system volume, while in the full CMF, including quarks, such a dependence is observed. A study of the $BS$-correlation in systems of different sizes would therefore be interesting to determine the appearance of deconfined quarks.

\section{Conclusion}

The effect of a finite system volume have been investigated within the Chiral Mean Field model which incorporates a transition between hadronic and quark degrees of freedom and is able to describe lattice QCD results in the thermal limit. In particular, the pressure, energy density, specific heat at constant volume, and ratios of conserved charge susceptibilities and correlations were studied at zero chemical potentials. The effect of finite volume is incorporated by introducing a lower momentum cut-off in the integrals of the particle momenta. Some of the quantities studied, such as the speed of sound $c_s^2$ show a significant dependence on the finite volume. As expected, the chiral crossover transition becomes smoother as the system volume becomes smaller, as indicated by the shallower minimum in the speed of sound. The mass spectrum of hadrons along with excluded volume governs the relative abundance of hadrons at a particular system volume through a nontrivial dependence. The system size dependence up to the crossover is dominated by the lightest bosonic degrees of freedom -- the pions and kaons -- and the ratios of susceptibilities indicate that the crossover transition is influenced to some extent by the finite system volume. The dependence on system volume is more significant as one considers a smaller system volume, comparable to the region of homogeneity in heavy-ion collisions. 
This dependence of the ratios of susceptibilities on the system volume is of great relevance if one tries to extract system properties from measured susceptibility ratios in experiments that suffer from short lifetimes and small system sizes. 
The observed finite-volume effects are different from those due to finite eigenvolumes in the excluded volume model, where the largest effect comes from hadrons which have large hard-core volumes, i.e. the baryons. 
It is found that the largest effect is due to the bosonic nature of the pions which makes them most sensitive to the low momentum cutoff. 
The effect can be expected to be stronger at finite chemical potentials. 
We also observe a notable suppression due to finite volume in strangeness related cross-susceptibilities, $\chi_{11}^{QS}$ and $\chi_{11}^{BS}$, which is due to the appearance of deconfined quarks.

This dependence of the ratios of susceptibilities on the system volume is of great relevance if one tries to extract system properties, like the freeze out temperature from measured susceptibility ratios in experiments that suffer from short lifetimes and small system sizes (see e.g. \cite{Alba:2014eba}). Similarly extracting the speed of sound from flow observables \cite{Pratt:2015zsa,OmanaKuttan:2022aml} may have to be corrected for finite size effects in the speed of sound. 
Finally, one should also keep in mind that in heavy ion collisions different observables are often presented as function of the centrality, i.e. system size. 
This means that comparing experimental measurements of conserved charge susceptibilities in heavy ion collisions with lattice QCD data would require corrections for the finite size (on top of corrections for global charge conservation) and should be done with much greater care.

\begin{acknowledgments}
SP thanks the Department of Science and Technology, India for support. 
AM acknowledges the Stern-Gerlach Postdoctoral fellowship of the Stiftung
Polytechnische Gesellschaft. HS gratefully acknowledges the Judah M. Eisenberg Laureatus Professorship of the Fachbereich Physik and the Walter Greiner Gesellschaft. The authors appreciate support by
the Alexander von Humboldt (AvH) foundation and the
BMBF through a Research Group Linkage program.
The computational resources for this project were provided by the Center for Scientific Computing of the GU Frankfurt and the Goethe-HLR.
\end{acknowledgments}

\bibliography{main}

\end{document}